\documentclass[%
 reprint,
 twocolumn,
superscriptaddress,
 amsmath,amssymb,
 aps,
prl,
]{revtex4-1}

\usepackage{graphicx}
\usepackage{dcolumn}
\usepackage{bm}
\usepackage{xcolor}

\newcommand{\p}{\partial}
\newcommand{\cV}{{\cal V}}
\newcommand{\cE}{{\cal E}}

\begin{document}


\title{Stable multi-ring and rotating solitons in two-dimensional spin-orbit coupled Bose-Einstein condensates with a radially-periodic potential}
 
\author{Yaroslav V. Kartashov}
\affiliation{%
ICFO-Institut de Ciencies Fotoniques, The Barcelona Institute of Science and Technology, 08860 Castelldefels (Barcelona), Spain
}%
\affiliation{%
Institute of Spectroscopy, Russian Academy of Sciences, Troitsk, Moscow, 108840, Russia
}%
 
\author{Dmitry A. Zezyulin}%
\affiliation{%
ITMO University, St.~Petersburg 197101, Russia
}%

\date{\today}

\begin{abstract}

We consider two-dimensional spin-orbit coupled atomic Bose-Einstein condensate in a radially-periodic potential. The system supports different types of stable self-sustained states including radially-symmetric vorticity-carrying modes with different topological charges in two spinor components that may have multiring profiles and at the same time remain remarkably stable for repulsive interactions. Solitons of the second type show persistent rotation with constant angular frequency. \textcolor{black}{They can be stable for both repulsive and  attractive  interatomic interactions.} Due to inequivalence between clockwise and counterclockwise rotation directions introduced by spin-orbit coupling, the properties of such solitons strongly differ for positive and negative rotation frequencies.   \textcolor{black}{Collision of solitons located in the same or different rings is accompanied by change of the rotation frequency that depends on the phase difference between colliding solitons.} 
\end{abstract}


\maketitle

Nonlinear wave phenomena in atomic Bose--Einstein condensates (BECs) attract considerable attention \cite{becgeneral1,becgeneral2,fetter2009}. Depending on the sign of the interatomic interactions one can observe the formation in BEC of bright or dark solitons. Their  
properties critically depend on the dimensionality of the condensate, since multidimensional states in BECs with attractive and repulsive interatomic interactions may be prone, respectively, to collapse or various snaking instabilities. \textcolor{black}{Bright solitons can be obtained in a Bose-Fermi mixture of degenerate gases, even if interatomic interaction in the Bose component are repulsive \cite{fermi,fermiexp}. Another} powerful approach to stabilization of multidimensional states in BEC relies on the external potentials, including periodic ones \cite{beclattice1,beclattice2,beclattice3}. Besides conventional states that do not change upon evolution \cite{beclattice4}, such potentials, when they are radially-symmetric \cite{radial1,radial2,radial3}, 
support stable solitons exhibiting persistent rotation. The properties of such solitons in single-component BECs 
do not depend on the rotation direction.

This situation may change dramatically in spin-orbit coupled two-component condensates (SO-BECs) which are 
attracting steadily growing interest. SO-BECs, 
representing a mixture of different states of the same atomic species, were recently used for demonstration of coupling between pseudospin degree of freedom and spatial structure of the condensate \cite{socbec1,socbec2,socbec3}. SO-BECs offer a versatile platform for investigation of the nonlinear phenomena in the presence of synthetic fields \cite{gauge1} and gauge potentials \cite{gauge2}, see review \cite{gauge3}. SO coupling notably modifies dispersion of the system \cite{stripe1,stripe2}, it may break Galilean invariance \cite{socbec2,galilean1}, and substantially impacts properties of one- \cite{free1dim1,free1dim2} and  multidimensional \cite{free2dim,free3dim}
solitons in the free space.
Especially intriguing is the impact of SO coupling on BEC in the external potentials, where possible symmetries of self-sustained states and their evolution dynamics are determined by the symmetry of the potential. It was studied for solitons on a ring \cite{onring} and in harmonic trap \cite{harmonic}, in toroidal traps \cite{toroidal}, Bessel \cite{beslattice}, and periodic \cite{perlattice} lattices.

While it was shown for radially-symmetric potentials \cite{toroidal} that SO coupling notably enriches two-dimensional soliton families and leads to appearance of azimuthal density modulations, the most important and unexpected manifestation of this effect, consisting in breakup of equivalence of two 
rotation directions (clockwise and counterclockwise) for solitons, was not demonstrated in atomic BECs. This is also the case for literature \cite{socrotate} on trapped SO-BEC under rotation that studies condensate transformation for one sign of the rotation frequency. While this inequivalence has been encounteblack in polariton condensates in a circular geometry \cite{polaritons}, polaritons   represent an essentially non-equilibrium system, where dominating interactions are  repulsive, and where effective SO coupling has completely different physical origin (it stems from TE-TM splitting) and is relatively weak. Thus, the question arises, whether this phenomenon exists in \textit{conservative} atomic BECs where SO coupling is   \textit{considerable} and where interactions can be both \textit{repulsive} and \textit{attractive}.

Here we first introduce SO coupling into BEC in a radially-periodic potential and show that  in the repulsive case
it supports stable \textit{multiring} vortex solitons carrying different topological charges in two components. Such structures have never been obtained in SO-BECs before and are in clear contrast to previously encountered azimuthally modulated patterns. Second, we show that radially-periodic potentials support stable \textit{rotating multipole} states for repulsive nonlinearity and \textit{crescent-like fundamental} states for attractive nonlinearity, reminiscent of azimuthons \cite{azimuthon}, that feature unexpected dependence not only on modulus, but also on \textit{sign} of the rotation frequency clearly illustrating inequivalence of two azimuthal directions. Third, we study interactions of rotating solitons for attractive nonlinearity and show that they lead to change in rotation frequency that also depends on soliton phase difference.

Meanfield dynamics of two-dimensional BEC is described by the normalized Gross-Pitaevskii equations with Rashba SO coupling \textcolor{black}{\cite{supp}}:
\begin{eqnarray}
\label{eq:GPE}
i\partial_t \psi_\pm = -(1/2)(\partial_{x}^2 + \partial_{y}^2) \psi_\pm +V(r)\psi_\pm
\pm\beta(\partial_x \mp i\partial_y)\psi_\mp
\hspace{0mm}\nonumber\\
+ \sigma(|\psi_\pm|^2 +|\psi_\mp|^2)\psi_\pm. \hspace{5mm}
\end{eqnarray}
Here $(\psi_+,\psi_-)^\textrm{T}$ is the spinor macroscopic wavefunction; \textcolor{black}{$t$ and $x,y$ are dimensionless time and spatial coordinates, scaled to characteristic time $\hbar/{\cal E}_0$ and spatial scale $R_0$, respectively; ${\cal E}_0=\hbar^2/mR_0^2$ is characteristic energy; $m$ is the atomic mass;   $\beta$ characterizes the strength of SO coupling that can be considerable; $\sigma=\pm1$ corresponds to repulsive/attractive interactions; $V(r)=2V_0\cos^2(r)$ is the radially-periodic potential with depth $2V_0$ measured in units of ${\cal E}_0$} (hereafter $r$ and $\theta$ are the polar  radius and  angle); in what follows, we set $V_0=3$. The radially-periodic potential can be created using a cylindrical laser beam whose amplitude is modulated with a patterned mask (the conical diffraction of the beam with the waist diameter $\simeq 100$~$\mu$m will be negligible for the tightly-confined  disk-shaped condensate with thickness $\simeq2$~$\mu$m \cite{radial3}). SO coupling is created by laser beams which couple different states of $^{87}$Rb atoms (the case of repulsive interactions) or $^7$Li atoms (attractive interactions); its strength can be varied in a broad range depending on laser configurations \cite{stripe2}; see also \cite{socbec1,socbec2,socbec3, free1dim1} for detailed discussion on implementation of SO-BECs.

\begin{figure}
\begin{center}		 \includegraphics[width=0.99\columnwidth]{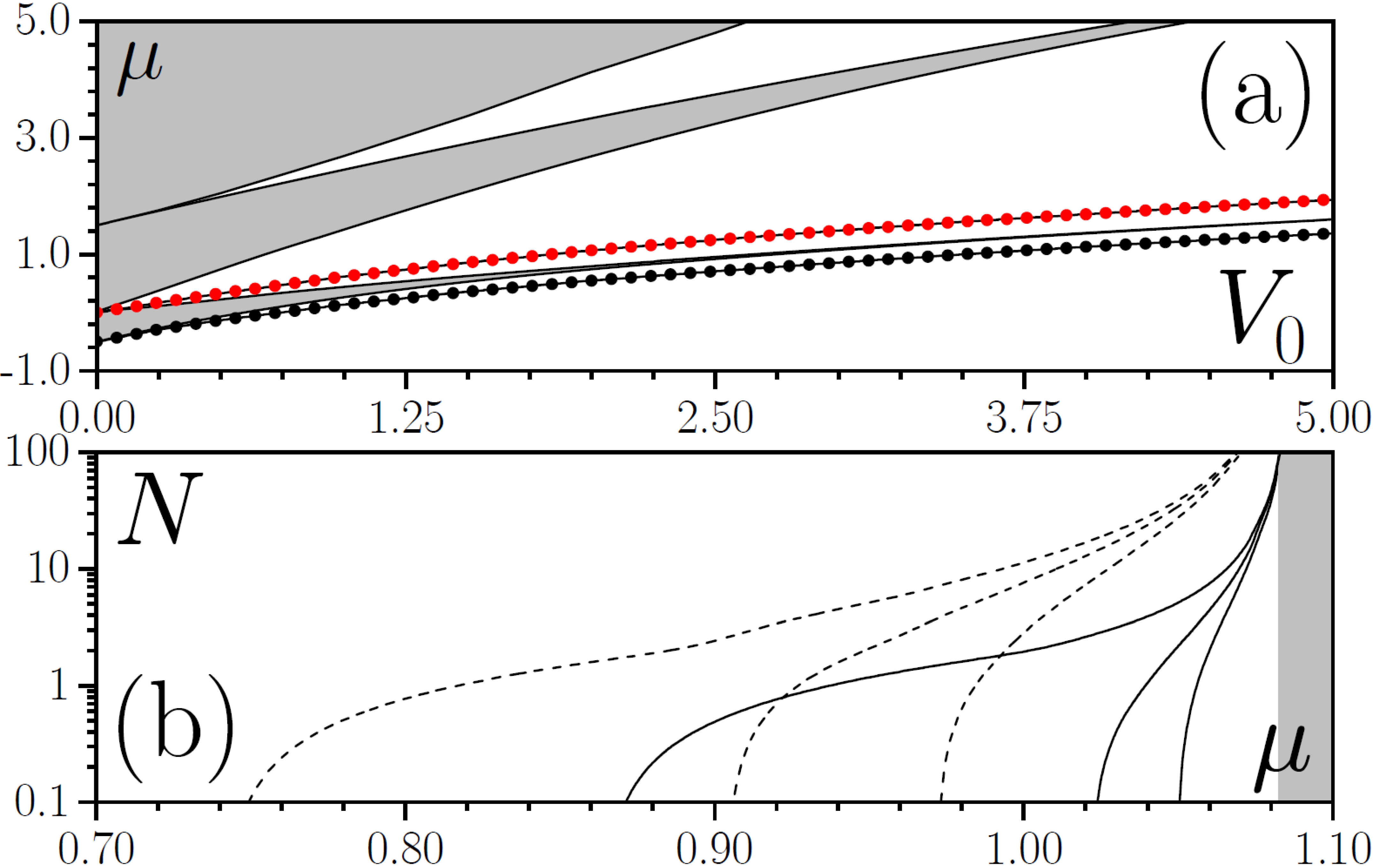}
\caption{(a) Bands (gray) and gaps (white) of radial potential and eigenvalues of linear modes (lines with circles) with $m_\pm=(-1,0)$ residing in the first potential minimum. (b) $N$ vs $\mu$ for simplest soliton families with $m_\pm=(-1,0)$ (solid lines) and $m_\pm=(-2,-1)$ (dashed lines) in the semi-infinite gap \textcolor{black}{at $\sigma=1$}. For each $m_\pm$ set three families are shown with density maxima in the first, second, and third minima of $V(r)$.}
\label{fig:figure1}
\end{center}
\end{figure}

\begin{figure}
\begin{center}		   \includegraphics[width=0.84\columnwidth]{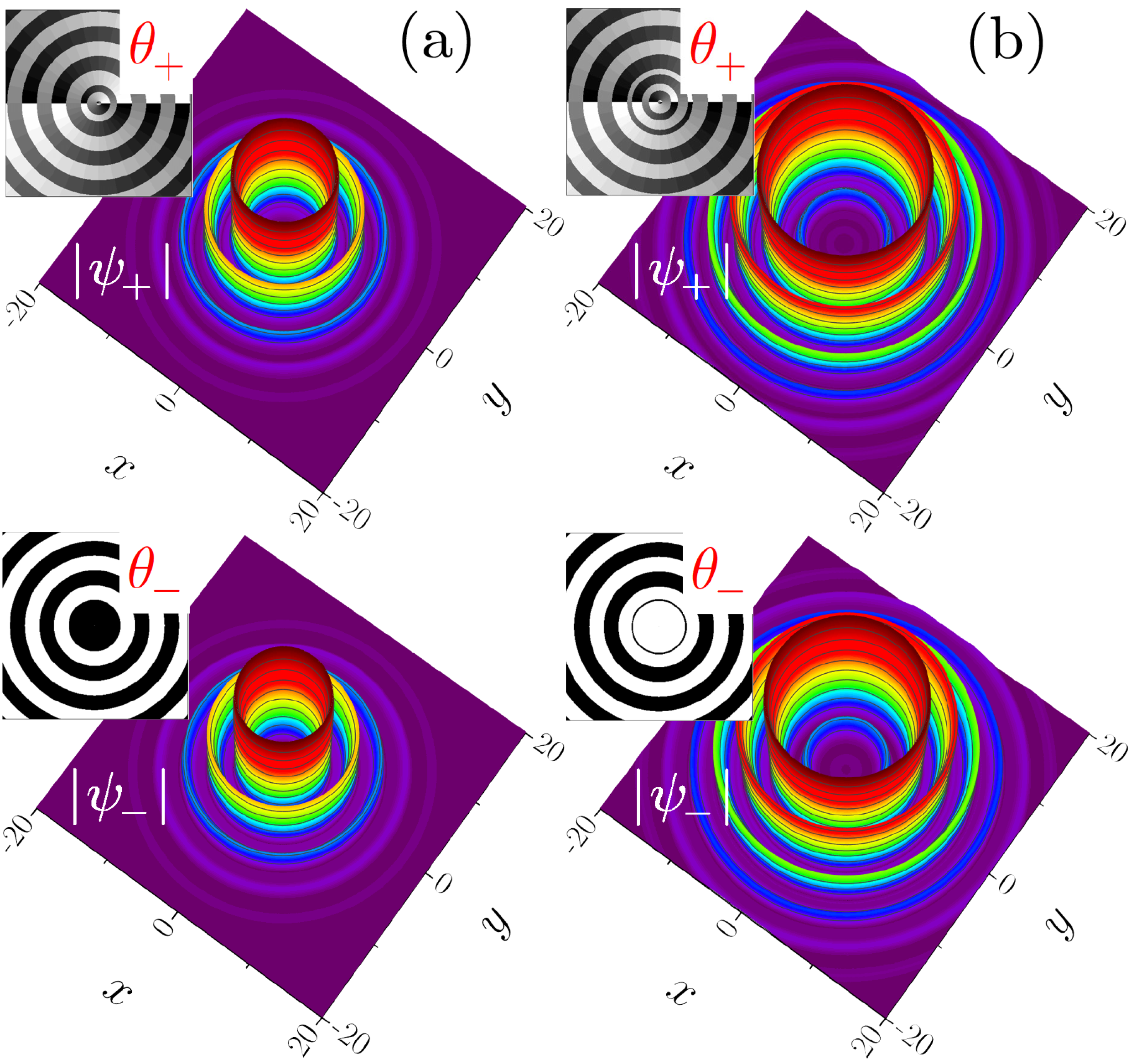}
\caption{Modulus and phase (insets) distributions in $m_\pm=(-1, 0)$ radially-symmetric solitons from different families, with $\mu=1.05$ (a) and $\mu=1.06$ (b) at $\beta=1$, \textcolor{black}{$\sigma=1$}.}
\label{fig:figure2}
\end{center}
\end{figure}

\begin{figure}
\begin{center}		   \includegraphics[width=0.99\columnwidth]{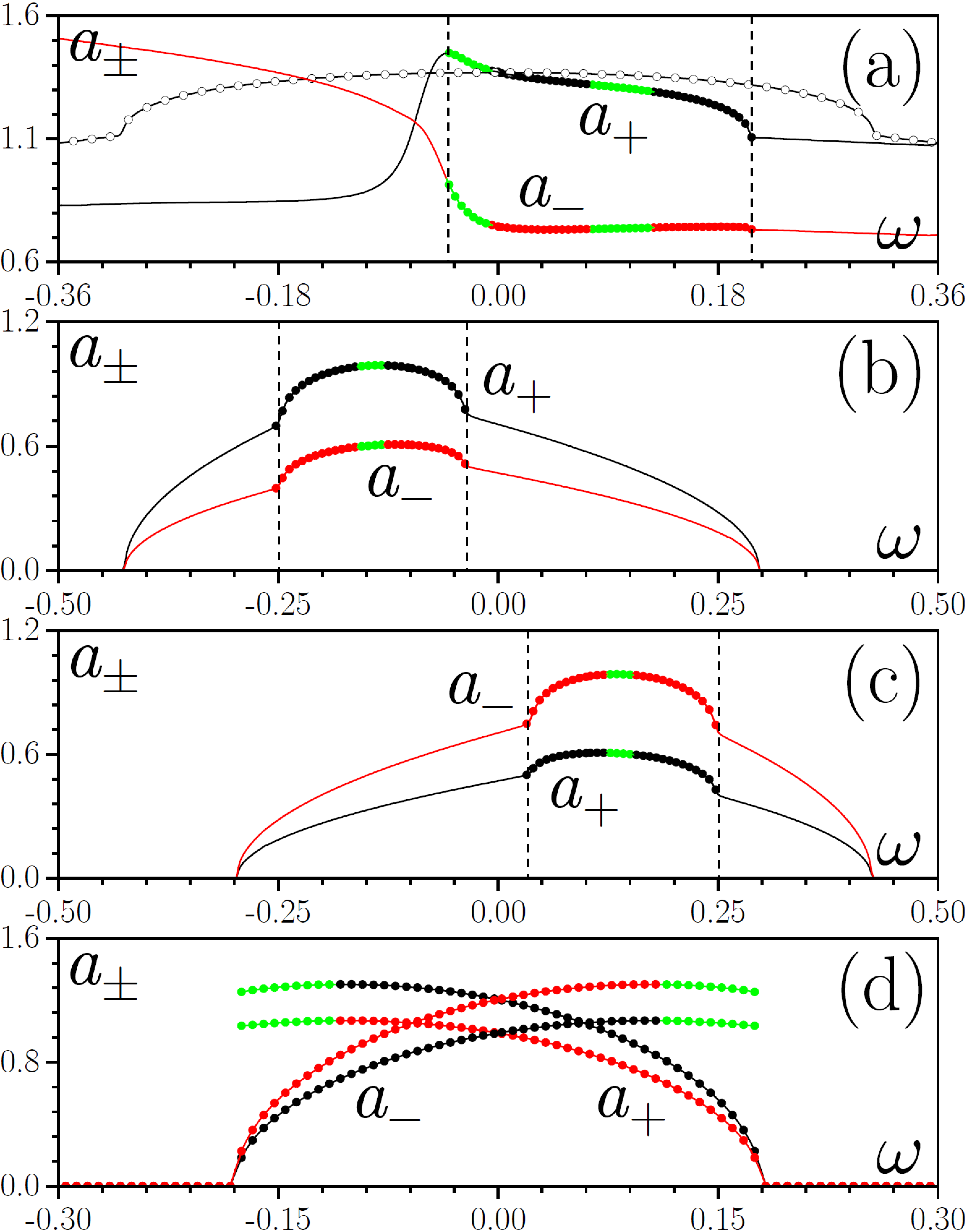}
\caption{Lines with solid circles show amplitudes $a_\pm$ of $\psi_\pm$  components in rotating dipole (a) and quadrupole (b),(c) solitons versus $\omega$ at $\mu = 3$, $\beta =0.5$, \textcolor{black}{$\sigma=1$}. Line with open circles in (a) shows $a_+(\omega)$  dependence at $\beta= 0$ . Dominating component is $\psi_+$ in (b) and $\psi_-$ in (c). Thin lines in (b),(c) correspond to radially-symmetric states. \textcolor{black}{(d) $a_\pm(\omega)$   in rotating fundamental solitons at $\mu=0.6$, $\beta=0.5$, $\sigma=-1$.} Green circles indicate unstable branches.}
\label{fig:figure3}
\end{center}
\end{figure}

The simplest states are radially-symmetric solitons $\psi_\pm = u_\pm(r)e^{-i\mu t + im_\pm \theta}$, where $m_\pm$ are the topological charges satisfying the condition $m_-=m_+ +1$ \textcolor{black}{that is due to the linear spin-orbit coupling preserving the total angular momentum}, $\mu$ is the chemical potential, and  $u_\pm$ are real-valued.
We search for localized solutions 
carrying finite norm $N=2\pi\int_0^\infty r(u_+^2 + u_-^2)dr$ (which is proportional to the total number of particles in the condensate). At $r\to\infty$ the effect of nonlinear terms becomes negligible, and 
the intervals of chemical potential, where   localized states can exist, are determined by the eigenvalue problem $\mu u_\pm = -({1}/{2})\,\partial_r^2 u_\pm + V(r) u_\pm \pm \beta\partial_r u_\mp$.
This problem is $\pi$-periodic and  features the bandgap spectrum shown in Fig.~\ref{fig:figure1}(a). Localized nonlinear modes --- \textcolor{black}{radial gap solitons} ---  exist  for $\mu$ values lying in the spectral gaps [white regions in Fig.~\ref{fig:figure1}(a)]. However, in contrast to usual gap solitons, nonlinear modes in the radially-periodic potential remain completely localized also in the small-amplitude limit, when the corresponding norm vanishes, $N\to 0$. This feature is readily visible from Fig.~\ref{fig:figure1}(b), where we plot 
several dependencies $N(\mu)$ for nonlinear modes with topological charges $m_\pm=(-1, 0)$ (solid curves) and $m_\pm=(-2, -1)$ (dashed curves) and density maxima located in  radial minima of the potential at $r=\pi/2$, $3\pi/2$, and $5\pi/2$. In the limit $N\to 0$, each of these soliton families bifurcates from appropriate localized linear mode with chemical potential $\mu$ from the gap. Eigenvalues of linear modes in the semi-infinite and first finite gaps from which the simplest solitons with density maximum at $r=\pi/2$ bifurcate are shown in Fig.~\ref{fig:figure1}(a) by lines with circles. In spite of the repulsive interactions, $\sigma=1$, families shown in Fig.~\ref{fig:figure1}(b) belong to the semi-infinite gap (which is possible due to the radial periodicity of the trap).
When $\mu$ approaches the edge of the gap,     vortex modes acquire well-pronounced multiring structure (see examples in Fig.~\ref{fig:figure2}). Standard linear stability analysis \cite{supp,Fetter01} and   direct integration of Eq.~(\ref{eq:GPE}) indicate  stability of all vortex states shown in Fig.~\ref{fig:figure1}(b) and Fig.~\ref{fig:figure2}, in spite of their complex multiring shapes.
Stable radially-symmetric vortex solitons can be found not only in the semi-infinite gap but also in finite spectral gaps, as shown in Figs.~\ref{fig:figure3} and \ref{fig:figure5}.

Now we turn to rotating states without radial symmetry. They are sought as $\psi_\pm=u_\pm(x',y')e^{-i (\mu \pm \omega/2) t}$ in the rotating frame $x'=x \textrm{cos}(\omega t)+y\textrm{sin}(\omega t)$, $y'=y\textrm{cos}(\omega t)-x \textrm{sin}(\omega t)$, where complex functions $u_\pm$ solve   (we further omit primes):
\begin{eqnarray}
\label{eq:roteq}
\mu u_\pm = -(1/2)(\partial_{x}^2 + \partial_{y}^2) u_\pm +V(r) u_\pm 
\pm\beta(\partial_x \mp i\partial_y)u_\mp
\hspace{0mm}\nonumber\\[1.0mm]
+\sigma (|u_\pm|^2 +|u_\mp|^2)u_\pm +i \omega (x\partial_y-y\partial_x)u_\pm \mp (\omega/2)u_\pm. \hspace{5mm}
\end{eqnarray}
Rotation with frequency $\omega$ results in  penultimate Coriolis term in Eq.~(\ref{eq:roteq}), while the last term originates from  the assumed form of time-dependence in $\psi_\pm$ that is required to eliminate time-dependence in SO coupling term in the rotating  frame. Equation~(\ref{eq:roteq}) admits a variety of rotating solitons residing in different radial minima of the potential. \textcolor{black}{We start with repulsive nonlinearity $(\sigma=1)$} and consider simplest solitons from the first minimum at $r=\pi/2$ with chemical potentials $\mu$ from the first finite gap. \textcolor{black}{At $\beta,\omega=0$ they represent dipole (two out-of-phase spots in $u_+$ or $u_-$ component) and quadrupole (four spots in $u_+$ or $u_-$ with $\pi$ phase jumps) solitons}. 
At $\beta=0$ increasing or decreasing rotation frequency smoothly transforms multipole states into radially-symmetric vortices. Dependence of amplitude $a_\pm=\textrm{max}|u_\pm(x,y)|$ of soliton components on $\omega$ is symmetric in this case, see line with open circles in Fig.~\ref{fig:figure3}(a) for dipole solitons. Thus at $\beta=0$ the properties of such solitons do not depend on the rotation direction. This picture changes qualitatively in the presence of SO coupling: the dependence $a_\pm(\omega)$ becomes 
asymmetric at $\beta \ne 0$. For dipole solitons with dominating $u_+$ component the entire domain of existence of rotating solitons shifts toward positive frequency values [lines with solid circles in Fig.~\ref{fig:figure3}(a) between two vertical dashed lines marking the border of the existence domain]. For dipole states with dominating $u_-$ component the existence domain shifts toward negative frequencies. The existence domains for rotating solitons with different dominating components are thus mirror-symmetric with respect to $\omega=0$, as illustrated for quadrupole solitons in Figs.~\ref{fig:figure3}(b,c). Inequivalence of azimuthal directions in SO-BEC 
illustrated by Fig.~\ref{fig:figure3} is one of the central results of this Letter. \textcolor{black}{For characteristic scale of $R_0=2~\mu \textrm{m}$,  dimensionless  frequency $\omega=0.25$ corresponds to rotation periods of $137~\textrm{ms}$ in $^{87}$Rb and $11~\textrm{ms}$ in $^7$Li condensate \cite{supp}, that is well below   condensate lifetime available in  the state-of-the-art
experiments.
}

\begin{figure*}
\begin{center}		   \includegraphics[width=0.99\textwidth]{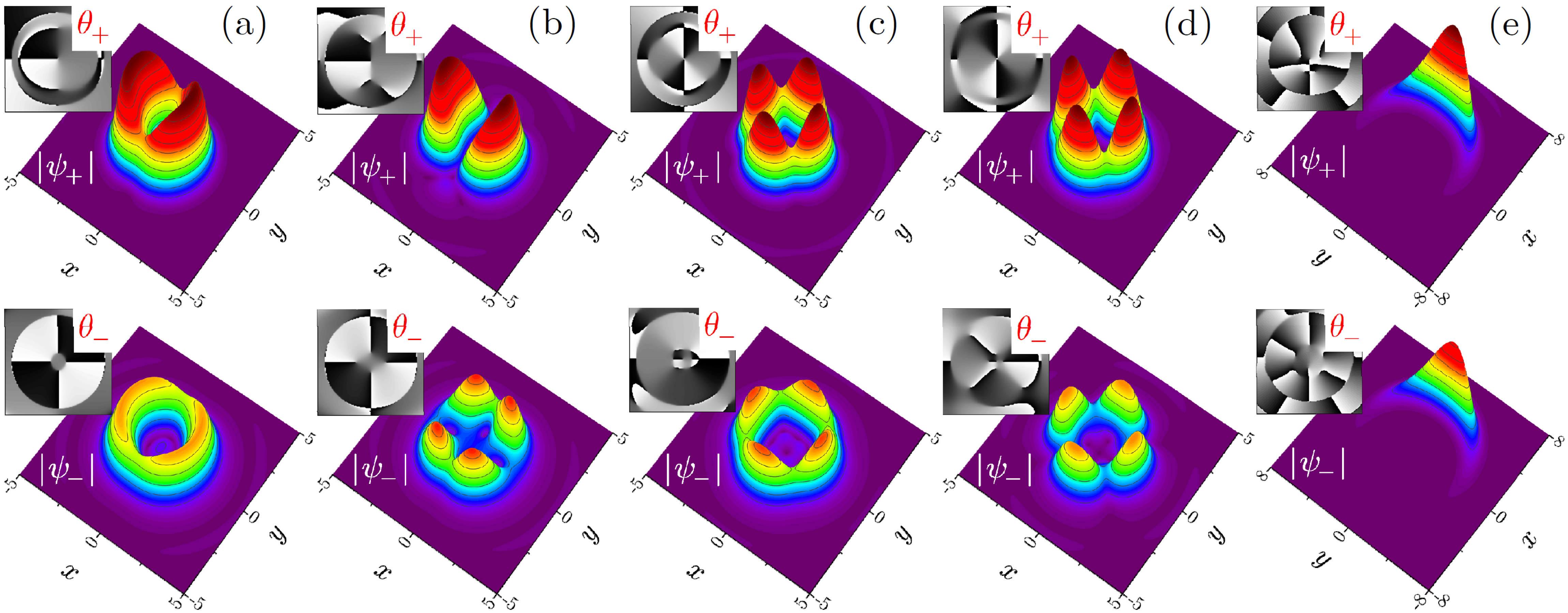}
\caption{Modulus and phase (insets) distributions in dipole solitons with (a)  $\omega=0.13$, (b) $\omega=-0.04$, and quadrupole solitons with (c) $\omega=-0.06$, (d) $\omega=- 0.22$ at $\mu=3$, $\beta= 0.5$, \textcolor{black}{$\sigma=1$, and fundamental solitons with (e) $\omega=0.15$ at $\mu=0.6$, $\beta=0.5$, $\sigma=-1$.} In all cases $\psi_+$ is a dominating component; $|\psi_+|$ and $|\psi_-|$ distributions are plotted with the same scale in each soliton.}
\label{fig:figure4}
\end{center}
\end{figure*}

Variation of rotation frequency 
causes notable shape transformations. Examples of modulus $|\psi_\pm|$ and phase $\theta_\pm$ distributions for different frequencies in dipole and quadrupole solitons 
are shown in Fig.~\ref{fig:figure4}(a-d). On the right edge of the existence domain in $\omega$ [Fig.~\ref{fig:figure4}(a)] such dipoles turn into $m_\pm=(-1,0)$ radially-symmetric vortices, while on the left edge [Fig.~\ref{fig:figure4}(b)] they   become 
strongly modulated and dynamically unstable. 
Quadrupole solitons  
transform into $m_\pm=(-2,-1)$ vortices  on the right edge [Fig.~\ref{fig:figure4}(c)] of the existence domain
and into $m_\pm=(+2,+3)$ vortices on its left edge [Fig.~\ref{fig:figure4}(d)]. \textcolor{black}{Transformation of phase distribution upon variation of $\omega$ resembles topological charge flipping \cite{flipping}}. 

Families of radially-symmetric states into which rotating solitons transform can be further continued in $\omega$ as shown by thin lines in Fig.~\ref{fig:figure3}. At fixed $\omega$ and $\sigma=1$ all rotating solitons (lines with circles) bifurcate with increase of chemical potential $\mu$ from radially-symmetric solitons (thin lines), which emanate from corresponding linear modes, as shown in Fig.~\ref{fig:figure5}. The charge of the state, from which bifurcation occurs, is determined by $\omega$: it is $(-1,0)$ in Fig.~\ref{fig:figure5}(a) and $(-2,-1)$ in Fig.~\ref{fig:figure5}(b). When $\mu$ increases and approaches the border of the first gap,   rotating modes develop multiring structure and eventually delocalize.

\begin{figure}
\begin{center}		   \includegraphics[width=0.99\columnwidth]{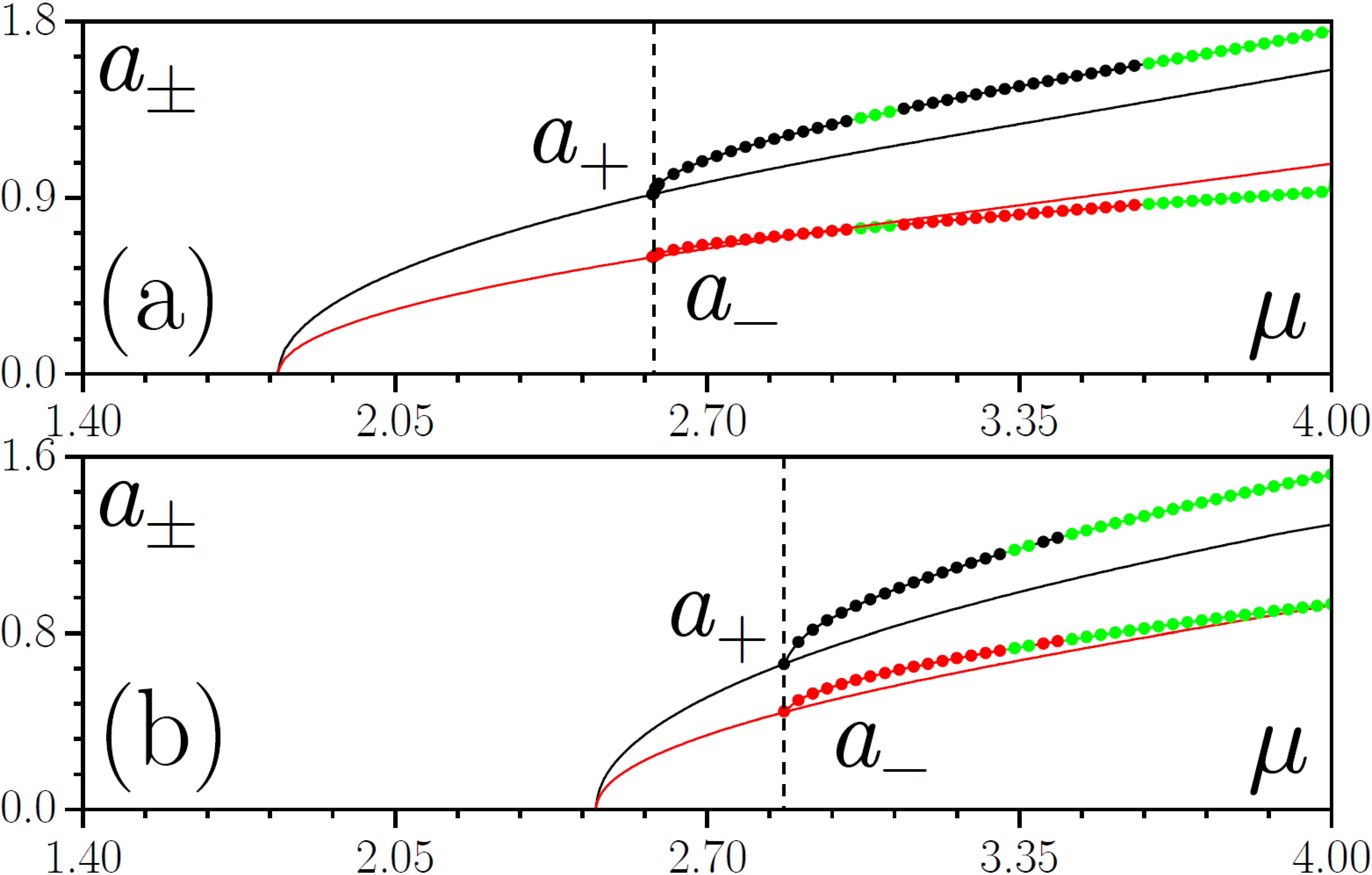}
\caption{Lines with circles: $a_\pm(\mu)$ dependencies for (a) rotating dipole soliton with $\omega =0.12$ and (b) quadrupole soliton with $\omega =-0.06$ at $\beta =0.5$, \textcolor{black}{$\sigma=1$}. Thin lines: $a_\pm(\mu)$ for radially-symmetric $m_\pm = (-1, 0)$ (a) and $m_\pm =(-2,-1)$ (b) states. Green circles indicate unstable branches.}
\label{fig:figure5}
\end{center}
\end{figure}

\begin{figure}
\begin{center}		   \includegraphics[width=0.99\columnwidth]{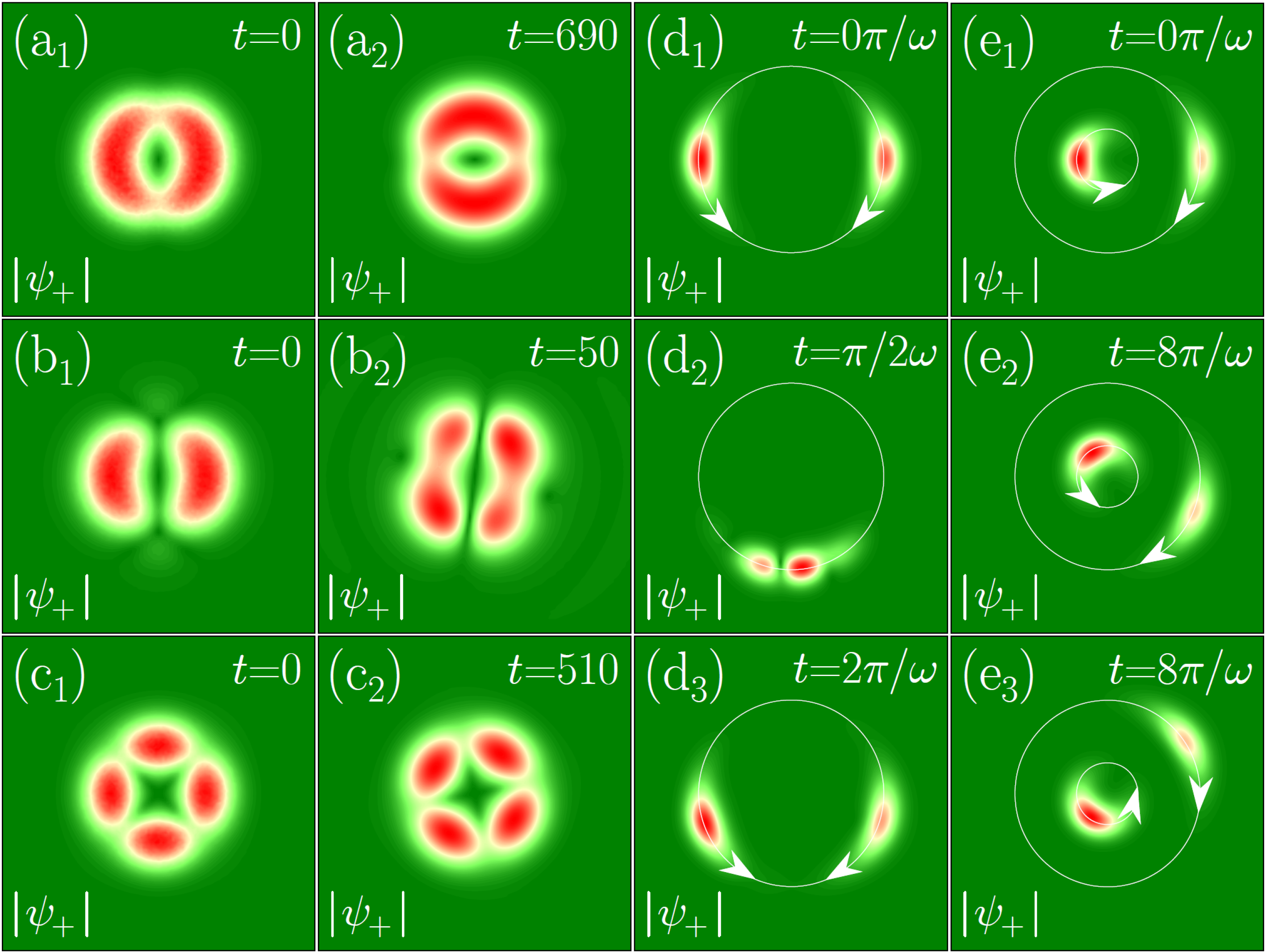}
\caption{Stable (a,c)  and unstable (b) evolution of rotating dipole and quadrupole solitons at (a) $\omega =0.13$, (b) $\omega=-0.04$, (c) $\omega=-0.22$ and $\mu=3$, $\beta =0.5$, $\sigma=1$. Interaction of two fundamental solitons \textcolor{black}{in the same (d$_1$-d$_3$) and in different (e$_1$-e$_3$) rings at $\mu=0.6$, $\beta=0.5$, $\sigma=-1$. Solitons are in-phase in (d$_1$-d$_3$) and (e$_2$) and out-of-phase in (e$_3$) and have opposite rotation frequencies $\omega=\pm1$.}}
\label{fig:figure6}
\end{center}
\end{figure}

In contrast to polariton condensates, both dipole and quadrupole rotating solitons in repulsive SO-BECs are dynamically stable in wide parameter regions even for $\beta$ values comparable to 1. \textcolor{black}{Stability was also tested by modeling the evolution of slightly perturbed states up to huge times $t \sim 10^4$ in Eq.~(\ref{eq:GPE})}. Instability domains are indicated by green circles in Figs.~\ref{fig:figure3} and \ref{fig:figure5}, while black/red circles correspond to stable branches. Rotating solitons are always stable in the parameter domains adjacent to bifurcation points from radially-symmetric solitons. Examples of evolutions with  stable and unstable rotations are given in Figs.~\ref{fig:figure6}(a,b,c).

\textcolor{black}{SO-BEC with attractive nonlinearity $(\sigma=-1)$ supports simpler fundamental rotating solitons 
with unusual crescent-like shapes, see Fig.~\ref{fig:figure4}(e). Such solitons exist in semi-infinite gap and \textcolor{black}{resemble whispering-gallery modes \cite{gallery}}. The dependencies of amplitudes 
$a_\pm$ of such solitons are also strongly asymmetric in rotation frequency $\omega$ [Fig.~\ref{fig:figure3}(d)]: at one border of the existence domain in $\omega$ soliton gradually expands along the entire ring where it resides (soliton is stable in broad domain adjacent to this border), while on another border one observes development of unstable multiring structure. 
Since such solitons are better localized than their counterparts in repulsive condensate, one can study their collision in the same [Fig.~\ref{fig:figure6}(d$_1$-d$_3$)] or in different [Fig.~\ref{fig:figure6}(e$_1$-e$_3$)] rings. Taking two solitons with opposite rotation frequencies $\omega$ and equal norms ($\mu$ values), one unexpectedly finds that \textcolor{black}{after collision [Fig.~\ref{fig:figure6}(d$_2$)]} both solitons \textit{accelerate} [compare input in Fig.~\ref{fig:figure6}(d$_1$) with output in Fig.~\ref{fig:figure6}(d$_3$) after $t=2\pi/\omega$], i.e., collision in this system changes rotation frequencies. Moreover, the variation in $\omega$ depends on phase difference between colliding solitons. Thus, when two solitons with opposite frequencies at $t=0$ collide in different rings [Fig.~\ref{fig:figure6}(e$_1$-e$_3$)] the outer (inner) soliton accelerates (decelerates) upon consecutive collisions for in-phase solitons [Fig.~\ref{fig:figure6}(e$_2$)], while for out-of-phase states this tendency reverses [Fig.~\ref{fig:figure6}(e$_3$)] leading to 
different final density distributions}.

Summarizing,  we demonstrated that SO coupling in trapped BEC with repulsive or attractive interaction breaks the equivalence of two  rotation directions. \textcolor{black}{Rotating solitons feature strongly asymmetric existence domains in rotation frequency and feature  non-conventional collisional behaviors involving  their acceleration/deceleration determined by the phase difference between solitons.}

\begin{acknowledgments}
	\textit{Acknowledgments}. The research of  D.A.Z. was  supported by megaGrant No.~14.Y26.31.0015 of the Ministry of Education and Science of Russian Federation and by Government of Russian Federation (Grant 08-08).
\end{acknowledgments}

\pagebreak

\pagebreak

\widetext

\setcounter{equation}{0}
\setcounter{figure}{0}
\setcounter{table}{0}
\setcounter{page}{1}
\makeatletter
\renewcommand{\theequation}{S\arabic{equation}}
\renewcommand{\thefigure}{S\arabic{figure}}
\renewcommand{\bibnumfmt}[1]{[S#1]}
\renewcommand{\citenumfont}[1]{S#1}

\begin{center}
	\textbf{\large Supplementary Material for \textit{Stable multi-ring and rotating solitons in two-dimensional spin-orbit coupled Bose-Einstein condensates with a radially-periodic potential}}
\end{center}
\begin{center}
Yaroslav V. Kartashov$^{1,2}$ and Dmitry A. Zezyulin$^3$
\end{center}


\begin{center}
$^1\,$\textit{ICFO-Institut de Ciencies Fotoniques, The Barcelona Institute of Science and Technology, 08860 Castelldefels (Barcelona), Spain}
\end{center}
\begin{center}
$^2\,$\textit{Institute of Spectroscopy, Russian Academy of Sciences, Troitsk, Moscow, 108840, Russia}
\end{center}
\begin{center}
$^3\,$\textit{ITMO University, St.~Petersburg 197101, Russia}
\end{center}

\subsection{Physical units}

Assuming tight harmonic confinement along the $Z$ axis, we start with a two-dimensional Gross-Pitaevskii equation for a disk-shaped condensate in the horizontal $(X, Y)$ plane (see e.g. \cite{supp1,supp2}):
\begin{eqnarray}
i\hbar \frac{\p\Psi_\pm}{\p T} = -\frac{\hbar^2}{2m}\left(\frac{\p^2\ }{\p X^2}+\frac{\p^2\ }{\p Y^2}\right)\Psi_\pm + 2\cV_0\cos^2(R/R_0)\Psi_\pm \pm \frac{\gamma\hbar^2k}{m}\left(\frac{\p\ }{\p X}\mp i\frac{\p\ }{\p Y}\right)\Psi_\mp +U(|\Psi_\pm|^2+|\Psi_\mp|^2)\Psi_\pm.
\label{eq:SMGPE}
\end{eqnarray}
Here $T$ is time; $X,Y$ are the spatial coordinates, $R^2=X^2+Y^2$; $2{\cal V}_0$ is the potential depth, $R_0$ is the characteristic spatial scale ($\pi R_0$ determines diameter of the ring, where first minimum of the potential is achieved); $m$ is the atomic mass; $k$ is the wavenumber of lasers creating spin-orbit coupling; $\gamma$ is the dimensionless coefficient depending on the particular laser scheme creating spin-orbit coupling \cite{supp2,supp3}. We consider the case when the $s$-wave scattering lengths between atoms in the two hyperfine states are equal. Then the interactions between atoms can be described by a single coefficient $U$, which for a disk-shaped condensate with dimensional wavefunction $\Psi_\pm$ normalized as $\iint_{-\infty}^\infty (|\Psi_+|^2+|\Psi_-|^2)dXdY=1$ has the form (see e.g. \cite{supp4})
\begin{equation}
    U = 2aN\sqrt{\frac{2\pi\hbar^3\omega_Z}{m}},
\end{equation}
where $a$ is the $s$-wave scattering length, $N$ is the total number of atoms, and $\omega_Z$ is the frequency of tight confinement along the $Z$ axis. Next, we introduce dimensionless spatial variables
\begin{equation}
x = X/R_0, \quad y = Y/R_0, \quad r = R/R_0,
\end{equation}
as well as characteristic energy ${\cal E}_0$ and dimensionless time as
\begin{equation}
\cE _0 = \hbar^2/m R_0^2, \quad t = (\cE_0/\hbar) T,
\end{equation}
and obtain Eqs.~(1) of the main text
\begin{eqnarray}
i\partial_t \psi_\pm = -(1/2)(\partial_{x}^2 + \partial_{y}^2) \psi_\pm +V(r)\psi_\pm
\pm\beta(\partial_x \mp i\partial_y)\psi_\mp + \sigma(|\psi_\pm|^2 +|\psi_\mp|^2)\psi_\pm,
\end{eqnarray}
where dimensionless potential depth $V_0 = {\cal V}_0/{\cal E}_0$ is used; the strength of spin-orbit coupling $\beta = {\gamma kR_0}$; the parameter $\sigma = \textrm{sign}\, a$; and dimensionless spinor wavefunction $\psi_{\pm}= |U/\cE_0|^{(1/2)}\Psi_{\pm}$ is introduced. For characteristic spatial scale of $R_0=2~\mu$m  (that gives diameter $6.28~\mu$m of the first circular minimum of potential), one obtains characteristic time scale of 5.45~ms (for $^{87}$Rb) or 0.44~ms (for $^7$Li). Thus, for typical dimensionless rotation frequency $\omega=0.25$ of the states from Figs.~3,~4,~6 of the main text, one rotation occurs on $\sim 137~\textrm{ms}$ for $^{87}$Rb and $\sim 11~\textrm{ms}$ for $^7$Li, i.e., the rotations times are below experimentally available lifetime of the condensate.


\subsection{Equations for linear stability analysis and numerical methods}

In order to address linear stability of radially-symmetric vortex modes $u_\pm(r)$, we use the standard substitution for slightly perturbed stationary solution:
\begin{eqnarray}
\label{eq:SMpert}
    \psi_\pm(x,y, t) = e^{-i\mu t + im_\pm \theta}[u_\pm(r)  +  w_{\pm,q}(r)e^{i\lambda t + iq\theta}+  z^*_{\pm,q}(r)e^{-i\lambda^* t - iq\theta}],
\end{eqnarray}
where $w_{\pm,q}$, $z_{\pm,q}$ are small radial perturbations, and integer $q = 0, 1, 2, \ldots$ is the azimuthal perturbation index. Spectrum of $\lambda$ determines the growth rate of an eventual instability (the instability corresponds to $\lambda$ with nonzero imaginary part). Substituting (\ref{eq:SMpert}) into dimensionless GPE and retaining only the terms linear in $w_{\pm, q}$ and $z_{\pm, q}$, after  straightforward calculations, for each azimuthal index $q$,  we arrive  to the linear eigenvalue problem  $L_q {\bf Z}_q = \lambda {\bf Z}_q$, where $\textbf{Z}_{q}=(w_{+,q}, z_{+,q}, w_{-,q}, z_{-,q})^\textrm{T}$, and differential operator $L_{q}$ has the form

\begin{equation*}
\left(
\begin{array}{cccc}
     M(m_++q) -2\sigma u_+^2 - \sigma u_-^2&
     -\sigma u_+^2&
     -\sigma u_+u_- - B(m_++q)&
     -\sigma u_+u_-
     \\[3mm]
     \sigma u_+^2
     & 
     -M(m_+-q) + 2\sigma u_+^2 + \sigma u_-^2&
     \sigma u_+ u_-&
     \sigma u_+u_- + B(-m_+-q)\\[3mm]
     -\sigma u_+u_- - B(-m_+ - q)&
     -\sigma u_+u_-&
     M(m_-+q) - 2\sigma u_-^2 - \sigma u_+^2&
     -\sigma u_-^2\\[3mm]
     \sigma u_+ u_-&
     \sigma u_+ u_- - B(q-m_+)&
     \sigma u_-^2&
     -M(m_--q) + 2\sigma u_-^2 + \sigma u_+^2
\end{array}
\right),
\end{equation*}
where we have introduced the following notations
\begin{eqnarray}
M(q)  =  \frac{1}{2}\left(\frac{\partial^2\,}{\partial r^2} + \frac{1}{r}\frac{\partial\,}{\partial r} - \frac{q^2}{r^2}\right) + \mu -V(r), \qquad 
B(q) = \beta\left(\frac{\partial\,}{\partial r} + \frac{q}{r}\right).
\end{eqnarray}
The obtained linear eigenvalue problem can be solved for various $q$ values using standard numerical finite-difference methods. In our work, we have considered the azimuthal indexes in the range  $q=0,1,\ldots 10$. 

Nonlinear stability of radially-symmetric and rotating states was also analyzed by means of direct integration of the time-dependent Gross-Pitaevskii equation (1) from the main text. The initial conditions were chosen in the form of slightly perturbed (by random or regular perturbations) exact soliton states. The time-dependent equation was solved with a split-step Fourier method, where the spatial differential operators were approximated using the fast Fourier transforms.

\end{document}